\documentclass[runningheads]{llncs}
\usepackage{amsmath}
\usepackage[T1]{fontenc}
\usepackage{graphicx}
\usepackage{cite}
\usepackage{algorithm}
\usepackage{algpseudocode}
\usepackage{amsmath,amssymb,amsfonts}
\usepackage{textcomp}
\usepackage{xcolor}
\usepackage{subcaption}
\usepackage{upgreek}

\newcommand{\squeezeup}{\vspace{-2.5mm}}

\usepackage{graphicx}
\begin{document}
\title{Parallel Approaches to Accelerate Bayesian Decision Trees}
%
%
\author{Efthyvoulos Drousiotis\inst{1} \and
Paul G, Spirakis\inst{2}\and
Simon Maskell\inst{1}}
\authorrunning{E Drousiotis et al.}
%

\institute{Department of Electrical Engineering and Electronics, University of Liverpool, Liverpool L69 3GJ, UK;
\email{\{E.Drousiotis, S.Maskell\}@liverpool.ac.uk} \and
Department of Computer Science, University of Liverpool, Liverpool L69 3BX, UK\\
\email{P.Spirakis@liverpool.ac.uk}\\
 }
\maketitle              
\squeezeup
\squeezeup
\squeezeup
\begin{abstract}

Markov Chain Monte Carlo (MCMC) is a well-established family of algorithms primarily used in Bayesian statistics to sample from a target distribution when direct sampling is challenging. Existing work on Bayesian decision trees uses MCMC. Unforunately, this can be slow, especially when considering large volumes of data. It is hard to parallelise the accept-reject component of the MCMC. None-the-less, we propose two methods for exploiting parallelism in the MCMC: in the first, we replace the MCMC with another numerical Bayesian approach, the Sequential Monte Carlo (SMC) sampler, which has the appealing property that it is an inherently parallel algorithm; in the second, we consider data partitioning.  Both methods use multi-core processing with a High-Performance Computing (HPC) resource. We test the two methods in various study settings to determine which method is the most beneficial for each test case. Experiments show that data partitioning has limited utility in the settings we consider and that the use of the SMC sampler can improve run-time (compared to the sequential implementation) by up to a factor of 343.

\keywords{Parallel algorithms  \and Machine Learning \and Bayesian Decision Tree}
\end{abstract}
\section{Introduction, Recent Work, and Our Contribution} \label{sec:relatedwork}

Markov Chain Monte Carlo(MCMC) can characterise a distribution without knowing the analytic form of that distribution. MCMC achieves this by using samples to characterise the distribution. The sampling operation is Markovian such that the new sample depends only on the previous one (and not any other samples before that previous one). MCMC is an extremely effective tool for  Bayesian inference because it can handle a wide variety of posterior distributions and specifically those distributions  which would be difficult or impossible to work with via analytic examination. MCMC has been used to solve Bayesian inference problems spanning biology \cite{valderrama2019mcmc}, forensics \cite{taylor2014interpreting}, education \cite{drousiotis2021early}, and chemistry \cite{dumont2021quantification}, among other areas.

As will be explained in more detail in section 3.1, MCMC operates by proposing samples and then accepting these samples with a probability that is calculated such that the MCMC generates samples from the distribution of interest. Since the next step of the MCMC depends on the current sample, it is non-trivial for a single MCMC chain to be processed contemporaneously by several processing elements. Byrd \cite{byrd2008reducing} proposed a method to parallelise a single MCMC chain, where backup samples are calculated in separate threads of execution in anticipation of the possibility that the current move is not accepted. This is wasteful since it will often be the case that the current move is accepted (such that the backup samples are not needed). However, when the acceptance rate is relatively low, there will be increased utility from considering such backup samples. Of course, there will be a trade-off between having many concurrent threads generating such backup samples and them all being used. More generally, single-chain parallelisation can quickly become problematic as the efficiency of the parallelisation is not guaranteed. 

Another way of improving the run-time of MCMC is to reduce the convergence rate by requiring fewer iterations. Metropolis-Coupled MCMC\cite{altekar2004parallel}, $(MC)^{3}$, runs multiple MCMC chains at the same time. One chain is treated as ``cold'' and targets the posterior of interest, while the other chains are treated as ``hot'', and attempt to sample from tempered versions of the posterior such that proposed samples are more likely to be accepted. The idea is that the space will be explored more quickly by the ``hot'' chains than by the ``cold'' chains: the ``hot'' chains find it easier to make what appear myopically to be disadvantageous moves across the parameter space. Results indicate improvements in run-time when more chains (and cores) were considered.

Several recent studies \cite{drousiotis2021capturing,jakaite2010bayesian,mohammadi2019continuous,vijayarangam2021novel} have considered MCMC in the context of Bayesian decision trees and as a replacement for traditional tree-based algorithms when tackling machine learning tasks. We are also aware that there are occassions (eg see \cite{kruijver2021estimating}) when researchers are eager to use MCMC Decision Trees, but it is not feasible or efficient for their research, forcing them to use alternative methods: the prohibitively expensive likelihood evaluations are frequently sufficiently time-consuming that these algorithms cannot be practically applied to complex models with large-scale datasets. This paper aims to investigate alternative mechanisms for achieving improved run-time in the context of Bayesian decision trees, and to explore how we can best apply such algorithms to large datasets with either small or big feature spaces. Our focus is on applications where run-time is critical. 

Existing papers consider a range of problem sizes. For example, \cite{jakaite2010bayesian} uses a dataset of 200 data, while \cite{mohammadi2019continuous} implemented some benchmarks results using a synthetic dataset of 300 data. There are also some comparison studies using Bayesian Trees; for example, \cite{vijayarangam2021novel} used 303 data points to train their algorithm, while \cite{drousiotis2021capturing, drousiotis2022balancing} compared different algorithms, including Bayesian Additive Regression Trees (BART), with a dataset of 2000 data points.

In this paper, we describe two methods with the aim of  achieving a faster execution time for Bayesian decision trees. Our contributions are then:
\begin{itemize}
\item We describe Sampling Using Multinomial Distribution (SUMD), a novel algorithm for performing inference in the context of Bayesian decision trees that uses a Sequential Monte Carlo (SMC) sampler to generate samples.
\item We describe Data Partitioning (DP), which, for transparency, we see as pragmatic and do not perceive to be novel and which spreads the data among the available cores, evaluating each tree faster than a conventional implementation of MCMC.
\item  We consider a diverse range of datasets to assess which of these approaches is the most suitable way of parallelising Bayesian decision trees, and identify the effective number of cores that should be used, given the type and the length of the dataset being processed.
\end{itemize}

The remainder of this paper is organised as follows: Section 2 presents the statistical model used in Bayesian Decision trees; Section 3 then describes the methods we compare; Section 4 presents the experimental results,  and Section 5 concludes the paper. 

\section{Bayesian Decision Trees}\label{Section3}

A decision tree operates by descending a tree. The process of outputting a classification probability for a given datum starts at a root node. At each non-leaf node, a decision as to which child node to progress to is made based on the datum and the parameters of the node. This process continues until a leaf node is reached. At the leaf node, a node-specific and datum-independent classification output is generated.

Our model describes the conditional distribution of a value for $Y$ given the corresponding values for $x$, where $x$ is a vector of predictors 
and $Y$ the corresponding values that we predict. We define the tree to be $T$ such that the function of the non-leaf nodes is to (implicitly) identify a region, $A$, of values for $x$ for which $p\left(Y|x\in A\right)$ can be approximated as independent of the specific value of $x\in A$, ie as $p(Y|x)\approx p(Y|\phi_j,x\in A)$. This model is called a probabilistic classification tree, according to the quantitative response $Y$.

For a given tree, $T$, we define the depth of the tree to be $d(T)$, the set of  leaf nodes to be $L(T)$ and the set of non-leaf nodes to be $\bar{L}(T)$. The tree, $T$, is then parameterised by: the set of features for all non-leaf nodes, $k_{\bar{L}(T)}$;  the vector of corresponding thresholds, $c_{\bar{L}(T)}$; the parameters, $\phi_{L(T)}$, of the conditional probabilities associated with the leaf nodes. This is such that the parameters of the tree are $\theta(T)=[k_{\bar{L}(T)},c_{\bar{L}(T)},\phi_{L(T)}]$ and $\theta(T)_j$ are the parameters associated with the $j$th node of the tree, where:
\begin{align}
\theta(T)_j = \left\{\begin{array}{ll}
\left[k_j,c_j\right] &j\in \bar{L}(T)\\
\phi_j& j\in L(T).\end{array}\right.
\end{align}

Given a dataset comprising $N$ data, $Y_{1:N}$ and corresponding features, $x_{1:N}$, and since decision trees are specified by $T$ and $\theta(T)$, a Bayesian analysis of the problem proceeds by specifying
a prior probability distribution, $p(\theta(T) ,T)$ and associated likelihood, $p(Y_{1:N}|T,\theta(T),x_{1:N})$. Because $\theta(T)$ defines the parametric model for $T$, it will usually be convenient to adopt the following structure for the joint probability distribution of $N$ data, $Y_{1:N}$, and the $N$ corresponding vectors of predictors, $x_{1:N}$: 
\begin{align}\label{fullformula}
 p(Y_{1:N},T,\theta(T)|x_{1:N}) =& p(Y_{1:N}|T,\theta(T),x_{1:N})p(\theta(T),T)\\
 =&p(Y_{1:N}|T,\theta(T),x_{1:N})p(\theta(T)|T)p(T)
\end{align}
which we note is proportional to the posterior, $p(T,\theta(T)|Y_{1:N},x_{1:N})$, and where we assume
\begin{align}
     p(Y_{1:N}|T,\theta(T),x_{1:N}) &= \prod_{i = 1}^{N} p(Y_i|x_i,T,\theta(T))   \label{labels probabiliteis}\\
    p(\theta(T)|T) &= \prod_{j\in T}p(\theta(T)_j|T) \\&= \prod_{j\in T} p(k_j|T)p(c_j|k_j,T)\label{features and thresholds}\\
    p(T) &= \frac{a}{(1+d(T))^\beta}\label{prior}
\end{align}

Equation \ref{labels probabiliteis} describes the product of the probabilities of every data point, $Y_i$, being classified correctly given the datum's features, $x_i$, the tree structure~$T$, and the features/thresholds, $\theta(T)$, associated with each node of the tree.
At the $j$th nod, equation~\ref{features and thresholds} describes the product of possibilities of picking the $k_j$th feature and corresponding threshold, $c_j$, given the tree structure, $T$.
Equation \ref{prior} is used as the prior for the tree $T$. This formula is recommended by \cite{chipman2010bart} and three parameters specify this prior: the depth of the tree, $d(T)$; the parameter, $a$, which acts as a normalisation constant; the parameter, ${\beta > 0}$, which specifies how many leaf nodes are probable, with larger values of $\beta$ reducing the expected number of leaf nodes. $\beta$ is crucial as this is the penalizing feature of our probabilistic tree which prevents an algorithm that uses this prior from over-fitting and allows convergence to occur\cite{rovckova2019theory}. Changing $\beta$ allows us to change the prior probability associated with ``bushy'' trees, those whose leaf nodes do not vary too much in depth.

An exhaustive evaluation of equation \ref{fullformula} over all trees, $T$, will not be feasible, except in trivially small problems, because of the sheer number of possible trees. 

Despite this limitations, Bayesian algorithms can still be used to explore the posterior. Such algorithms simulate a chain sequence of trees, such as:

\begin{equation}\label{chain}
    T_0, T_1, T_2,....,T_n
\end{equation}
which converge in distribution to the posterior, which is itself proportional to the joint distribution,
$p(Y_{1:N}|T,\theta(T),x_{1:N})p(\theta(T)|T)p(T)$, specified in equation \ref{fullformula}. We choose to have a simulated sequence that gravitates toward regions of higher posterior probability. Such a simulation can be used to search for high-posterior probability trees stochastically. We now describe the details of  algorithms that achieve this and how they are can be implemented.

\subsection{Stochastic Processes on Trees}

To design algorithms that can search the space of trees stochastically, we first need to define a stochastic process for moving between trees. More specifically, we consider the following four kinds of move from one tree to another:
\begin{itemize}
\item Grow(G) : we sample one of the leaf nodes, $j\in L(T)$, and replace it with a new node with parameters, $k_j$ and a $c_j$, which we sample uniformly from their parameter ranges.
\item Prune(P) : we sample the $j$th node (uniformly) and make it a leaf.
\item Change(C) : we sample the $j$th node (uniformly) from the non-leaf nodes, $\bar{L}(T)$, and sample $k_j$ and a $c_j$ uniformly from their parameter ranges.
\item Swap(S) : we sample the $j_1$th and $j_2$th nodes uniformly, where $j_1\neq j_2$, and swap $k_{j_1}$ with $k_{j_2}$ and $c_{j_1}$ with $c_{j_2}$.
\end{itemize}

We note that there will be situations (eg when moving from a tree with a single node) when some moves cannot occur. We assume each `valid' move is equally likely and this then makes it possible to compute the probability of transition from one tree, $T$, to another, $T'$, which we denote $q\left(T',\theta(T')|T,\theta(T)\right)$.

\section{Methods}
\subsection{Conventional MCMC}\label{sec:MCMC}

One approach is to use a conventional application of Markov Chain Monte Carlo to decision trees, as found in~\cite{drousiotis2022single}.

More specifically, we begin with a tree, $T_1$ and then at the $i$th iteration, we propose a new tree by sampling $T'\sim q\left(T',\theta(T')|T_i,\theta(T_i)\right)$. We then accept the proposed tree by drawing $u\sim U([0,1])$ such that:
\begin{align}
T_{i+1} = \left\{\begin{array}{ll} T' & u\leq\alpha(T'|T)\\
T_i & u > \alpha(T'|T)\end{array}\right.
\end{align}
where we define the acceptance ratio, $\alpha(T',T)$ as:
\begin{align}
\alpha(T'|T) = \frac{p(Y_{1:N}|T,\theta(T),x_{1:N})}{p(Y_{1:N}|T',\theta(T'),x_{1:N})}\frac{q\left(T,\theta(T)|T',\theta(T')\right)}{q\left(T',\theta(T')|T,\theta(T)\right)}
\end{align}.

This process proceeds for $n$ iterations. We take the opportunity to highlight that this process is inherently sequential in its nature.

\subsection{Sampling Using Multinomial Distribution (SUMD)}

We now consider using an Sequential Monte Carlo (SMC) sampler\cite{del2006sequential} to tackle the problem of sampling the trees from the posterior. We begin with a $C$ trees (which we assume to be sampled from some initial distribution over trees with a target that is that initial prior such that the inital weights are all $\frac{1}{C}$), one for each of the $C$ cores that we have available: we denote the $c$th tree at the $i$th iteration, $T^{(c)}_i$. At each iteration and for each tree, we sample $T^{(c)}_i\sim q(T^{(c)}_i,\theta(T^{(c)}_i)|T^{(c)}_{i-1},\theta(T^{(c)}_{i-1}))$. This is such that the weights of the $i$th tree is $\alpha(T^{(c)}_i|T^{(c)}_{i-1})$. We then normalise the weights (such that they sum to unity) and implement multinomial resampling to replace the trees with trees to which we assign equal weights (of $\frac{1}{C}$). We take the opportunity to emphasise that, just like MCMC, an SMC sampler configured in this way will generate samples from the posterior. 

This method is summarised in  algorithm \ref{DecisionTreeMetropolisParallel}. Note that we have replaced the widely-used and hard to parallelise accept-reject process in MCMC with multinomial sampling and that we can evaluate $\alpha(T^{(c)}_i|T^{(c)}_{i-1})$ in parallel across the $C$ cores (we are also aware of techniques for implementing the multinomial sampling in parallel (eg see \cite{varsi2021log2,varsi2020fast}) but have not found this component of the algorithm to be a computational bottleneck for the application considered herein). Once the algorithm has converged, we keep the samples drawn from the multinomial distribution.

\begin{figure}[!htb]
\begin{algorithm}[H]
\caption{Sampling Using Multinomial Distribution(SUMD)}\label{DecisionTreeMetropolisParallel}

\scriptsize
\begin{algorithmic}
\State C = Number of available cores
\State Initialize vector $T = [t_0, ..., t_c]$ with initial samples
\State $i = 0$
\State iterations = 8000
\State $Burn \;in \;period = iterations \;/ \;number\;of\;cores /2$
\While{$i < (iterations /\; number\;of\;cores)$}
    
    \State Allocate each $T^{(c)}$ to a separate core
    \State Sample each $T^{(c)_i}\sim q(T^{(c)}_i,\theta(T^{(c)}_i)|T^{(c)}_{i-1},\theta(T^{(c)}_{i-1}))$ in parallel
    \State Calculate  $w^{(c)}_i=\alpha(T{^{(c)}_i}|T{^{(c)}}_{i-1})$ in parallel
    Normalise the weights, $w^{(1)}_i\ldots w^{(C)}_i$\
    \State Draw $C$ samples from a $multinomial\; distribution$ parameterised by the normalised weights
    \State Generate the new (resampled) values for $T^{(c)_i}$
    \If{$i \geq Burn \;in \;period$} 
    \State Store the values of $T^{(1)}_i\ldots T^{(C)}_i$
    \EndIf
\State $i++$
\EndWhile
\end{algorithmic}
\end{algorithm}
\end{figure}

\subsection{Data Partitioning (DP)}
This method implements the same algorithm as is described in section~\ref{sec:MCMC}. The only difference is that we split the dataset, $Y_{1:N}$, across the number of available cores. For example, if the number of available cores $C=6$, we split the dataset into $6$ parts of equal size (in terms of the number of entries on each core). Each core processes the data in parallel and calculates the leaf probabilities for each leaf node with each core then calculating a contribution the (single) calculation of equation \label{eq:label_probabiliteis}.

We note that the SUMC approach departs from the use of MCMC with the aim of increasing the number of trees considered simultaneously and thereby reducing the number of iterations needed. In contrast, this second DP approach spreads the data among the available cores to more rapidly evaluate the tree sampled at each iteration of the MCMC.

\section{Experimental Setup and Results}

\subsection{Datasets}
To demonstrate the run-time improvement we achieved through our proposed methods, we experiment on 6 publicly accessible\footnote{https://archive.ics.uci.edu/ml/index.php} datasets listed in Table 1. We assert that this is a diverse range of datasets, and spans datasets with a small number of data points and small feature spaces, up to large datasets with large feature spaces. In particular, the mean number of records used is 300,480, with the lowest having 1,599 records and the largest 999,999. The mean number of features considered is 35, with the lowest being 9 features and the largest 91. For the sake of making comparisons more straightforward, we have categorised our datasets (see the class column in Table 1) into 6 classes based on whether the data is Small (SD), Medium (MD) or Big (BD) and whether the set of features considered in Small (SF) or big (BF). Henceforth, we refer to the datasets based on this classification.

\begin{table}[]
\caption{Datasets Specifications}
\begin{center}
\begin{tabular}{llll}
\hline
Dataset Name & Records & Features & Class \\ \hline
Abalone(AB) & 4,177 & 9 & SD/SF \\
Diabetes 012 Health(DH) & 253,680 & 22 & MD/BF \\
One Hundred Plant Species(OHPS) & 1,599 & 65 & SD/BF \\
OULAD & 28,084 & 9 & MD/SF \\
Poker Hand(PH) & 999,999 & 11 & BD/SF \\
Year Prediction MSD(YPM) & 515,344 & 91 & BD/BF
\end{tabular}
\end{center}
\end{table}

This section presents the experimental results obtained using the proposed methods with a focus on  run-time and convergence diagnostics. The following results were obtained using a local HPC platform comprising twin SP 6138 processors (20 cores each running at 2GHz), with 384GB memory RAM. For the large datasets, we have used high-memory nodes, each with 1TB of RAM. We note that the run-times presented in this paper include  any communication overheads.  

For testing purposes and a fair comparison and evaluation, we use the same hyperparameters $a$ and $\beta$ for every contesting algorithm. Moreover, for every method, we set the number of iterations to $8000$ and the number of burn-in iterations to $4000$. 

A very crucial aspect when designing parallel algorithms as alternatives to sequential alternatives is to ensure that the results output by the parallel algorithms offer commensurate (if not identical) results to the sequential algorithm. Table \ref{Accuracy} shows that the sampled trees output from SUMD were indeed commensurate in classification accuracy to those output by the conventional MCMC approach. We do not compare the accuracy of DP with the conventional MCMC approach since the method is not altering the operation of the original algorithm, but simply altering its implementation. 

\begin{figure}[!h]
  \centering
  \begin{subfigure}[b]{0.45\linewidth}
    \includegraphics[width=\linewidth]{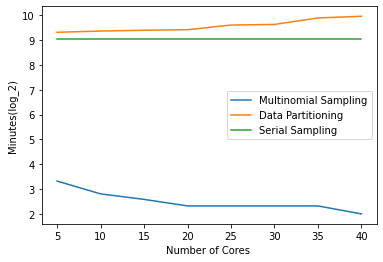}
    \caption{Small Feature Space}
  \end{subfigure}
  \begin{subfigure}[b]{0.45\linewidth}
    \includegraphics[width=\linewidth]{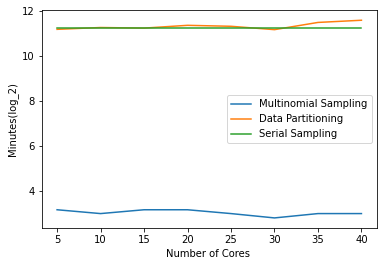}
    \caption{Big Feature Space}
  \end{subfigure}
  \caption{Run-time for Small datasets.}
  \label{Small_Run_Time}
\end{figure}

\begin{figure}[!h]
  \centering
  \begin{subfigure}[b]{0.450\linewidth}
    \includegraphics[width=\linewidth]{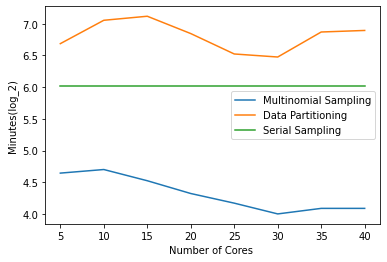}
    \caption{Small Feature Space}
  \end{subfigure}
  \begin{subfigure}[b]{0.45\linewidth}
    \includegraphics[width=\linewidth]{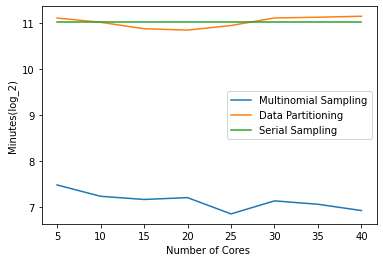}
    \caption{Big Feature Space}
  \end{subfigure}
  \caption{Run-time for Medium sized datasets.}
  \label{Medium_Run_Time}
\end{figure}

\begin{figure}[!h]
  \centering
  \begin{subfigure}[b]{0.45\linewidth}
    \includegraphics[width=\linewidth]{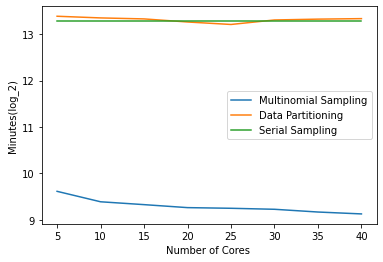}
    \caption{Small Feature Space}
  \end{subfigure}
  \begin{subfigure}[b]{0.45\linewidth}
    \includegraphics[width=\linewidth]{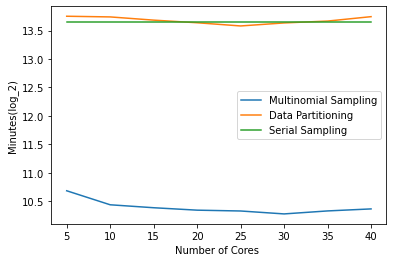}
    \caption{Big Feature Space}
  \end{subfigure}
  \caption{Run-time for Big datasets.}
  \label{Big_Run_Time}
\end{figure}

We tested each method with 10-Fold Cross Validation. Results indicate and validate our findings in section~\ref{sec:relatedwork} that running an MCMC Decision Tree on large datasets in a sequential implementation is very challenging when time is heavily constrained. More specifically, the run-time for BD/SF took 165~hours (approximately 7~days), and 213~hours (approximately 9~days) for BD/BF (see figures \ref{Small_Run_Time}, \ref{Medium_Run_Time} and\ref{Big_Run_Time}). Our results indicate that SUMD is much faster than the sequential implementation and DP in all cases considered (see Figures \ref{Small_Run_Time}, \ref{Medium_Run_Time} and \ref{Big_Run_Time}). In summary: SUMD is 132 times faster than the sequential implementation on SD/SF, 343 times faster on SD/MF, 4 times faster for MD/SF, 18 times faster for MD/BF, 19 times faster for BD/SF, and 10 times faster for BD/BF. 

We do note that SUMD is memory hungry: for the SUMD method, we need to evaluate many more trees concurrently (subject to the number of cores we are using) and each tree needs to be evaluated against all the data.

We anticipated DP would have resulted in greater improvements in run-time relative to the sequential implementation of MCMC. The underlying reason for this is that while we save time by having more processors, each handling less data, the communication overhead is significant when transferring data between the cores: for each datum, the MCMC needs access the probability vector for the corresponding leaf node and then calculate the probability of each data point being classified correctly. For this particular method, we also tried to partition the data once, when we produced the probabilities for each leaf node, to reduce the communication overhead. However, the resulting run-time was very close to that for our initial implementation, where we partitioned the data twice. Our conclusion is that, given the time spent communicating between the cores and the time needed to transfer the data to each core, there is very limited benefit stemming from parallelising MCMC decision tress by partitioning of the data.

Large datasets utilising more than 30 cores can only run in high memory nodes (with 1TB of memory RAM) when running entirely in parallel. Of course, if the nodes have less memory, it is possible to run with partial parallelism, where a core may wait for another core to continue its work. While the presented methods can run on a personal computer, it is not recommended: greater improvements to run-times and can be achieved using HPC. 

Figures \ref{Small_Run_Time}, \ref{Medium_Run_Time}, \ref{Big_Run_Time} make it possible to identify the recommended number of cores we should use in the context of each test case. Specifically, when the feature space is small, it is more beneficial, in terms of run-time, to use the maximum number of available cores we have. Experiments have shown that when the feature space is big, it is wiser to use 20 to 30 cores, as having more cores degrades run-time.

\begin{table}[]
\caption{Accuracy Comparison Between Sequential, and SUMD methods\label{Accuracy}}
\begin{center}
\begin{tabular}{|l|ll|}
\hline
\textbf{Dataset} & \multicolumn{2}{c|}{Accuracy} \\ \cline{2-3}
 & \multicolumn{1}{l|}{Sequential} & SUMD \\ \hline
SD/SF & \multicolumn{1}{l|}{22.48±0.2} & 22.48±0.2 \\ \hline
SD/BF & \multicolumn{1}{l|}{1.66±0.4} & 1.66±0.5 \\ \hline
MD/SF & \multicolumn{1}{l|}{63.23±0.1} & 63.23±0.1 \\ \hline
MD/BF & \multicolumn{1}{l|}{84.28±0.05} & 84.28±0.05 \\ \hline
BD/SF & \multicolumn{1}{l|}{26.56±0.07} & 26.56±0.1 \\ \hline
BD/BF & \multicolumn{1}{l|}{8.05±0.01} & 8.05±0.0.1 \\ \hline
\end{tabular}
\end{center}
\end{table}

\section{Conclusion}
Our comparison study has shown that by taking advantage of multi-core architectures, we can improve the run-time of Bayesian Decision Trees by up to $343$ times. According to our experimental results, our novel approach based on Sequential Monte Carlo samplers, SUMD, outperforms an approach that exploits data parallelism, with SUMD improving the run-time by an average of 88 times relative to a sequential Markov chain Monte Carlo implementation. Moreover, we have shown that the results obtained are very similar to the sequential implementation. 
We perceive that, by taking advantage of the improvements in modern processor designs, our methods help make Bayesian decision tree-based solutions more productive and increasingly applicable to a broader range of applications.

This study, though, has some limitations. First, to run in parallel and produce the speedups we described, the proposed methods require access to HPC nodes with substantial memory resources. Secondly, at present, the implementations we have are not yet as easy-to-use as existing (single-processor) libraries: we plan to address this in future work.

\bibliography{bib}

\end{document}